# Distinct scaling behaviors of giant electrocaloric cooling performance in low-dimensional organic, relaxor and anti-ferroelectrics

**Yuping Shi[1,2,3], Limin Huang[2,*], Ai Kah Soh[3], George J. Weng[4], Shuangyi Liu[5] & Simon A.T. Redfern[6,7]**

[1]Department of Mechanical and Aerospace Engineering, Hong Kong University of Science and Technology, Clear Water Bay, Kowloon, Hong Kong.

[2]Department of Chemistry, South University of Science and Technology of China, Shenzhen 518005, China.

[3]School of Engineering, Monash University Malaysia, Bandar Sunway 46150, Malaysia.

[4]Department of Mechanical and Aerospace Engineering, Rutgers University, New Brunswick, New Jersey 08903, USA.

[5]Chongqing Institute of Green & Intelligent Technology, Chinese Academy of Sciences, Chongqing 400714, China.

[6]Department of Earth Sciences, University of Cambridge, Downing Street, Cambridge CB2 3EQ, UK.

[7]HPSTAR, 1690 Cailun Rd, Pudong District, Shanghai 201203, China.

[*]Correspondence and requests for materials should be addressed to L.H. (email: huanglm@sustc.edu.cn)

## ABSTRACT

Electrocaloric (EC) materials show promise in eco-friendly solid-state refrigeration and integrable on-chip thermal management. While direct measurement of EC thin-films still remains challenging, a generic theoretical framework for quantifying the cooling properties of rich EC materials including normal-, relaxor-, organic- and anti-ferroelectrics is imperative for exploiting new flexible and room-temperature cooling alternatives. Here, we present a versatile theory that combines Master equation with Maxwell relations and analytically relates the macroscopic cooling responses in EC materials with the intrinsic diffuseness of phase transitions and correlation characteristics. Under increased electric fields, both EC entropy and adiabatic temperature changes increase quadratically initially, followed by further linear growth and eventual gradual saturation. The upper bound of entropy change ($\Delta S_{max}$) is limited by distinct correlation volumes ($V_{cr}$) and transition diffuseness. The linearity between $V_{cr}$ and the transition diffuseness is emphasized, while $\Delta S_{max}$=300 kJ/(K.m$^3$) is obtained for $Pb_{0.8}Ba_{0.2}ZrO_3$. The $\Delta S_{max}$ in antiferroelectric $Pb_{0.95}Zr_{0.05}TiO_3$, $Pb_{0.8}Ba_{0.2}ZrO_3$ and polymeric ferroelectrics scales proportionally with $V_{cr}^{-2.2}$, owing to the one-dimensional structural constraint on lattice-scale depolarization dynamics; whereas $\Delta S_{max}$ in relaxor and normal ferroelectrics scales as $\Delta S_{max} \sim V_{cr}^{-0.37}$, which tallies with a dipolar interaction exponent of 2/3 in EC materials and the well-proven fractional dimensionality of 2.5 for ferroelectric domain walls.

## Introduction

The electrocaloric effect, i.e. reversible changes in isothermal entropy or adiabatic temperature in polar materials achieved by application and removal of electric fields, has been extensively investigated in the context of next-generation techniques for more efficient and environmentally-friendly solid-state refrigeration and integrable on-chip coolers[1-4]. In addition to the pioneering discovery of giant EC responses in poly(vinylidene fluoridetrifluoroethylene) P(VDF-TrFE)-based polymer[5] ferroelectrics (FEs) and $PbZr_{0.95}Ti_{0.05}O_3$ antiferroelectric (AFE) thin-films[6], recent multiscale calculations[7,8] and indirect experiments[9-12] reliant on measuring the thermal dependence of FE hysteresis loops, have reported comparable EC cooling performances in a rich class of EC materials which also include normal and relaxor FEs as well as in a diverse range of low-dimensional structures such as nanotubes[13], nanowires[14] and nanocomposites[15]. Although direct EC tests based on *in-situ* measurement of the temperature changes accompanying adiabatic depolarizations have been implemented in FE films and multilayers (typically with microscale thickness)[16,17], direct experiments on solid-state cooling responses in low-dimensional EC structures still remains challenging due to their very small heat capacities, deviations from adiabatic conditions and the insufficient yet ultrafast heat transfer to thermal testing units[4,18]. Instead, development of a versatile theoretical framework capable of quantifying the interesting cooling properties in a broad range of EC materials is imperative and timely for the further enhancement of electro-thermal energy converting strength and in the search for new room-temperature refrigeration and microcooler alternatives.

When subjected to electric field ($E$) de-poling effects and temperature ($T$) variations, EC entropy changes ($\Delta S$) and adiabatic temperature changes ($\Delta T$) originate from partial reorientation of localized dipoles in organic FEs and from 1$^{st}$ or 2$^{nd}$ order phase transitions in AFEs, normal and relaxor FEs. Consequently, adiabatic depolarization dynamics in the rich variety of EC materials display varied transition diffusenesses and show a huge span in critical dimensions. The latter can range from lattice-scale dipole-dipole correlations to features associated with long-range FE domain structures. Furthermore, the cooling strength of EC effect (ECE) has distinct figures of merits among comprehensive electrically polarizable EC materials - the ECE in normal FEs becomes dramatically enhanced near any sharp phase transition and is seen predominantly around the Curie temperature ($T_c$); whereas, diffuse phase transitions in relaxor FEs dominated by the evolution of polar nano regions (PNRs) can show significant EC responses over a much broader temperature range[19,20], which often takes place far above $T_c$ and can substantially expand the scope and potential of the EC effect for solid-state and flexible refrigeration. When it comes to organic FEs, P(VDF-TrFE)-based copolymers usually behave like a normal FE; however, its terpolymers with chlorofluoroethylene (CFE) are display diffuse dielectric properties[18] with respect to both varying $E$ and $T$. These observations highlight the importance of considering both the overall phase transition features[21], as well as microscopic correlation characteristics and the lattice symmetry[22] of diversified EC materials when characterizing their cooling performances whether for academic or industrial motivations.

## Results

The Master equations are introduced here to describe the reorientation dynamics of localized depolarization, in which microscopic correlation characteristics and the lattice symmetry of EC materials are particularly taken into account. The overall polarization magnitude ($P$) and concomitant configurational entropy are treated as a spatially-averaged reflection throughout a sufficiently large ensemble of microscopic polar elements (MPEs), e.g. PNRs, ferroelectric domains or polarizable chemical chains, which correlate with varied strength through electronic or elastic interactions and over different length scales. In our framework, a mean characteristic volume of $V_{MPE}$ is particularly considered for an EC material or structure[19,20]. When $E$ is applied along the polarization direction or the axis of lattice symmetry, a universal activation parameter of both $E$ and $T$ is generalized as $u(E,T)=EP^{m}\mathrm{V}_{MPE}/(\Omega kT)$ where $P^m$, $\Omega$ and $k$ denote the maximum attainable polarization in the EC material, a symmetry factor and Boltzmann's constant, respectively. Subsequently, the solution of the established Master equations (see Methods) gives rise to a generic expression for the spatial ($u$) variation of the magnitude of overall polarization as $P=P^{m}\tanh(u)$, in which the symmetry factor is found to be 1 for normal and organic FEs and 3 for [111]-type depolarizations. The figure of merits of this symmetric polarization formulation with $E$ precludes the unwanted complexity of Joule heating induced by non-linear polarization hysteresis, thus offering an opportunity to improve the understanding of cyclically and directly-measured EC cooling performances.

In order to embrace the effects of phase transition diffuseness and the Curie temperature on EC cooling strength, the characteristic correlation volume may be related to the thermal diffuseness in the form of:[23]

$$\mathrm{V}_{MPE}(T)=\frac{\mathrm{V}_{cr}}{1+(T/T_{cr})^{n}} \tag{1}$$

where $n$ is referred as a diffuseness index to distinguish different types of EC phase transitions; $\mathrm{V}_{cr}$ is the saturated value of $\mathrm{V}_{MPE}$ at low temperatures; $T_{cr}$ denotes a critical temperature that can be the Curie temperature for normal FEs or the temperature where the maximum dielectric response occurs in continuously depolarized EC materials. The three-dimensional (3D) plot of equation (1) (Figure SI1) shows that increased diffuseness of $\mathrm{V}_{MPE}$ dispersion with normalized temperature $t = T/T_{cr}$ accompanies a decreased $n$. This indicates a rather smaller diffuseness index for relaxor FEs than that for sharp transitions in normal FEs. When the EC depolarizations are mediated by indivisible MPEs, say unit lattices or chemical chains, then $n$=0 is expected to ensure that $\mathrm{V}_{MPE}$ is constant, which is likely the case for polymeric EC materials. A direct correspondence of our equation (1) to the experimental correlation length data[24,25] in typical relaxor $Pb(Mg_{1/3}Nb_{2/3})O_3$ (PMN) is demonstrated in Figure 1(a). Since diffuse phase transitions shift the EC $\Delta T$ and $\Delta S$ peaks towards higher temperatures above $T_c$, the focal comparison in Figure 1(a) is made on the $t > 1$ region; the $n$ value of well-studied PMN single crystals is shown to lie between 2 to 3, implying a fractional PNR dynamics[26].

Upon successful establishment of the generic polarization expression with thermally diffuse correlation volumes, it is possible for us to elucidate $\Delta S$ and $\Delta T$ with the help of Maxwell relations: $dS/dE = \left(\partial P/\partial T\right)_E$ and $dT/dE = -\dfrac{T}{C_{\mathrm{v}}}\left(\partial P/dT\right)_E$, where $C_{\mathrm{v}}$ is the volumetric heat capacity of EC materials. Integrating the two Maxwell relations from zero field to $E$ yields

$$-\Delta S = \frac{\Omega k}{\mathrm{V}_{cr}}\left[1+\left(n+1\right)t^{n}\right]f_u\left(E,t\right) \quad \text{and} \quad \Delta T = \frac{\Omega k T_{cr}}{\mathrm{V}_{cr}C_{\mathrm{v}}}\left[t+\left(n+1\right)t^{n+1}\right]f_u\left(E,t\right) \tag{2}$$

where $f_u\left(E,t\right) = 2u - \ln\left(\dfrac{1+e^{2u}}{2}\right) - \dfrac{2u}{1+e^{2u}}$ governs the intrinsic EC responses of electric depolarizations with diffusionless characteristic correlation volumes, i.e. $n$=0. However, when the $\mathrm{V}_{MPE}$ is thermally diffuse in EC materials, according to equation (1), an EC enhancement function $f_n\left(t\right) = 1+\left(n+1\right)t^{n}$ and $t+\left(n+1\right)t^{n+1}$ immediately operates on both the $\Delta S$ and $\Delta T$. Noting the irrelevance of $f_n\left(t\right)$ with $E$ and the definition of ECE responsivity of $\Delta T/E$, a function of $f_u/u$ can describe the $E$-dependence of ECE responsivity as well. We find that equation (2) limits the upper bound of EC entropy change for the EC materials with non-diffuse $\mathrm{V}_{MPE}$ as $(-\Delta S)_{\max}=\ln(2)\Omega k/\mathrm{V}_{cr}$, which tallies with the $(-\Delta S)_{\max}$ result of thermodynamic and statistical mechanics[27]. Moreover, truncating the correlation length in nanocomposited or nanoconfined EC materials can enhance simultaneously the $\Delta T$ maxima and $(-\Delta S)_{\max}$, both of which are urgently required in realistic EC refrigeration applications.[1,18]

The functional dependences of $f_u\left(E,t\right)$ and ECE responsivity on a normalized $E$ are plotted in Figure 1(b), where three distinct changing trends are underlined for $f_u\left(E,t\right)$ at a specific $T$: an initial quadratic ($f_u = u^2/2$) increase, a further linear $\left(f_u \sim E\right)$ increment and gradual saturation at ultrahigh $E$. On the contrary, ECE responsivity is revealed as a non-monotonic function of $E$, but can still be approximated by a linear growth and then an exponential decay as $E$ increases from 0. The acquired quadratic increase of EC entropy and temperature change with $E$ successfully reproduces the $\Delta T$ changes with $E^2$ in dipole glasses[28] measured as early as in 1965 and the recently-proposed quadratic scaling law of $\Delta T$ in $PbZrO_3$ AFE[29]. It is also worth noting that both the proportional increase and the saturating trend of either $\Delta T$ or ($-\Delta S$) are frequently observed in pretty rich low-dimensional EC materials[30-34].

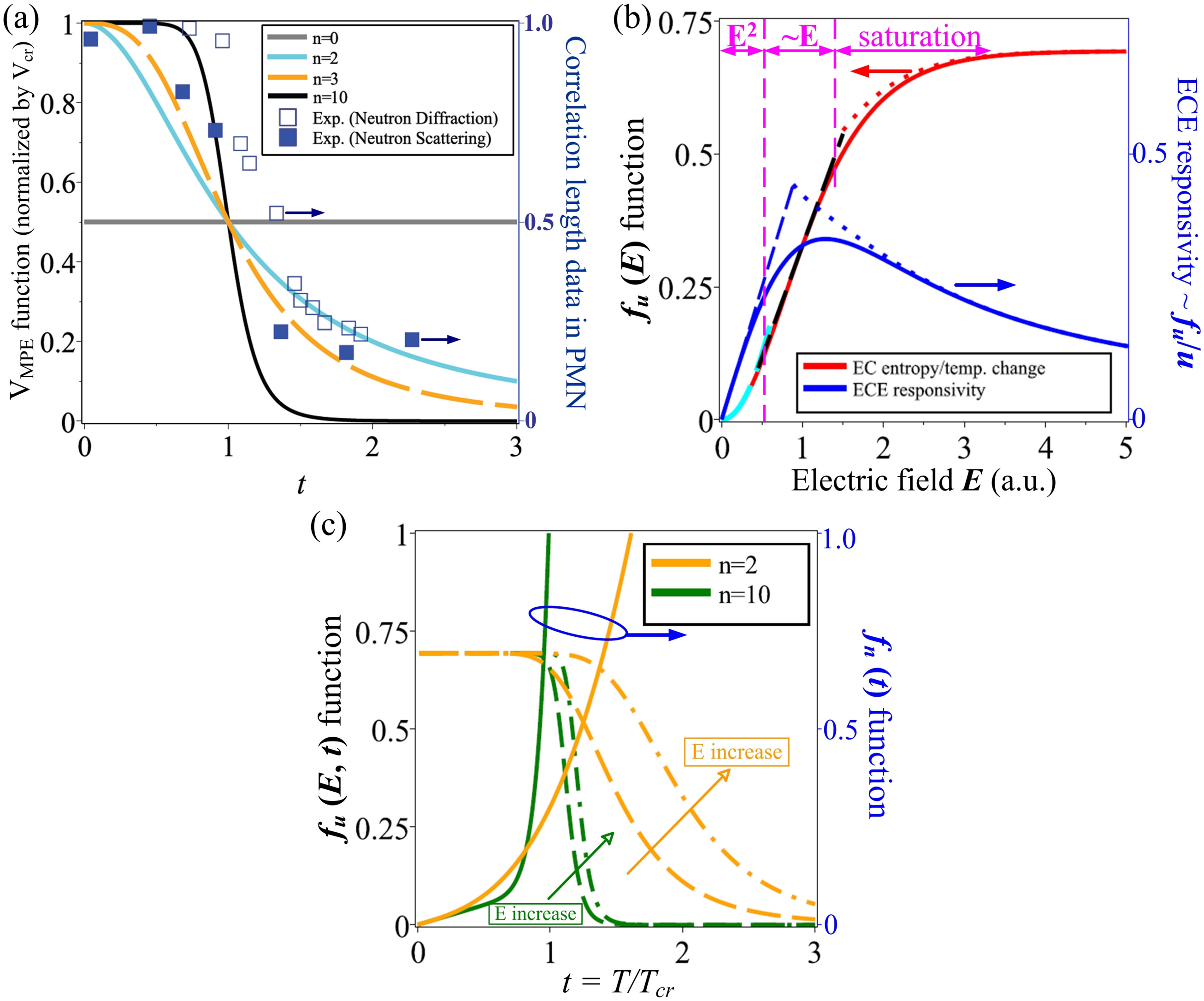

**Figure 1 |** (a) Direct comparison of the $V_{MPE}$ function with experimental correlation length data in PMN single crystals measured by neutron elastic diffuse scattering (NEDS) and neutron diffraction (ND) technique. (b) Dependence of EC entropy/temperature change (red curve) and ECE responsivity (blue curve) on normalized $E$ at a fixed temperature, as given by the $f_u(E,t)$ and $f_u/u$ function, respectively. (c) The $f_n(t)$ function for $\Delta T$ and $f_u(E,t)$ function at $n$=2 and $n$=10 vs. $t$, where the hollow arrows point to the increasing direction of $E$ and the higher $E$ used is ten times the lower $E$. The NEDS and ND correlation length data in (a) derived from refs. 24 and 25 is normalized by 65 nm and 20 nm, respectively; the $T_{cr}$ is chosen as 220 K for the NEDS data and 240 K for the ND data.

Figure 1(c) illustrates the $n$ and $E$ influence on the thermal evolution of the enhancement function for $\Delta T$ and the $f_u(E,t)$ function. Taking an EC material with $n$=10 for example, $f_u(E,t)$ exhibits abrupt change only over a narrow $T$ interval and slight above $T_{cr}$, which are also not very sensitive to $E$ increase even by as much as a factor of 10. This is in sharp contrast to the $n$=2 EC materials like PMN, where both $f_n(t)$ and $f_u(E,t)$ disperse over a wide $T$ range between $T_{cr}$ and $3T_{cr}$ and an amplified $E$ effectively shifts a specific value of $f_u(E,t)$ towards higher temperatures. It can be seen that these observations capture the main features of 1$^{st}$ order FE and 2$^{nd}$ order relaxor phase transitions, and therefore ensure equation (2) is feasible for quantifying the EC cooling responses from adiabatic electric depolarization in sharp and diffuse phase transitions.

We consider three categories of EC materials and low-dimensional structures based on a differentiation factor of $E_bT_{cr}$ (herein $E_b$ is dielectric breakdown field) - bulk FE, normal FE thin-films with long-range correlation and films of diffuse EC materials. These should possess a small, high and medium differentiation factors, respectively. Their three dimensional $\Delta T$ changes with $n$ and $t$ are calculated at first for a constant $V_{cr}$-dominated coefficient in equation (2) and are illustrated in Figures 2(a-c). As expected, $\Delta T$ in lower-$n$ FE materials has an expanded peak across a broad $T$ interval above $T_{cr}$; whereas the $\Delta T$ in high-$n$ EC materials maximize and decay steeply around $T_{cr}$ and likely has a second peak corresponding to the so-called dual EC peaks phenomenon[35] [Figures SI2(a-b)]. We particularly examine the case with a medium differentiation factor because it closely resembles typical currently-studied EC low-dimensional structures. An EC material with a lower $n$ is shown to have $\Delta T$ maxima at higher $T$ than that of higher-$n$ FEs. Based on an absence of $\Delta T$ peaks observed in PVDF-based polymer FEs below the melting point[5,15] and commonly observed $\Delta T$ peaks in normal FE thin films[7,11], projected $\Delta T$ 3D surfaces are given in Figures 2(l, m) by restricting $t$ between $0.5T_{cr}$ and $2T_{cr}$ setting $n < 1$ for relaxor organic FEs and $n > 4.5$ for normal FE ultrathin films, with $n$=1-4.5 for relaxor oxide EC materials. These $n$-assignments are marked in Figure 2(i), which shows the projected $\Delta T$ surface of the medium factor EC materials in Figure 2(d) that is computed by multiplying all the three functional terms in equation (2) and base on a proportionality between $n$ and $V_{cr}$. Accordingly, the overall ($-\Delta S$) surface and related projections under the same conditions are plotted in Figures 2(e, j, o). It is apparent that both EC $\Delta T$ and $\Delta S$ peak at $T_{cr}$ at the diffusionless phase transitions of normal FEs and that their maxima shift to higher temperatures by $0.4T_{cr}$ (i.e. over one hundred K since most EC materials have a $T_{cr}$ above 300 K) as $n$ decreases from 10 to 0 in diffusive depolarized EC materials.

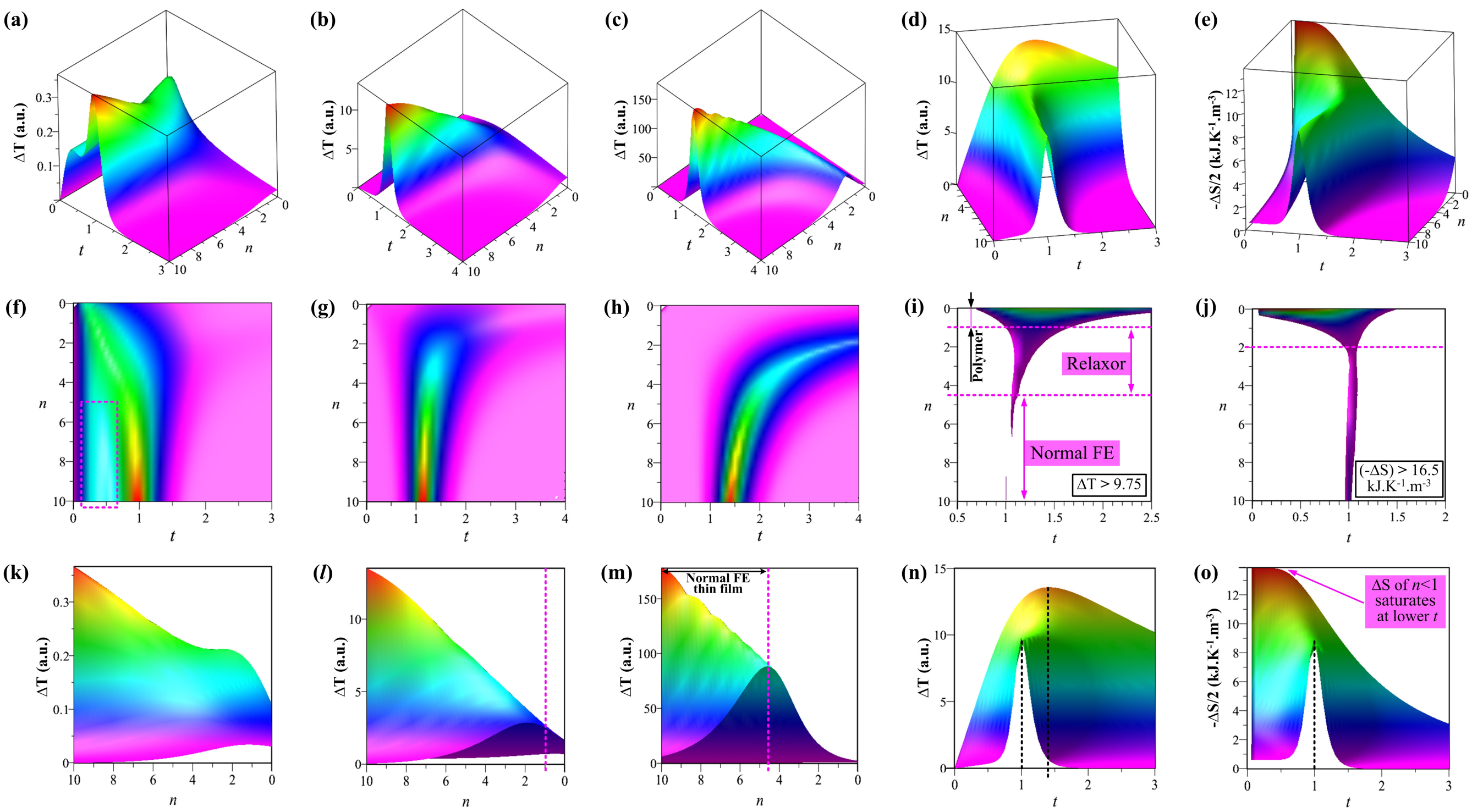

**Figure 2 | 3D plots of our calculated EC cooling performances vs. *t* and *n*.** Calculated ΔT surfaces of EC materials with a Small (a, f, k), Medium (b, g, l) and High (c, h, m) differentiation factor. The 3D surfaces of (d, i, n) ΔT and (e, j, o) (-ΔS) are calculated for the medium factor case using $V_{cr}=7.28(1+n)$. The panels in lower two rows are projections of the corresponding ones in the first row; (i) and (j) shows projected partial surface of $\Delta T \geq 9.75$ in (d) and $(-\Delta S) \geq 16.5$ in (e) onto the $n$ - $t$ plane, respectively. (k-m) are attained by projecting (a-c) for $t = 0.5$ - 2. The dashed box in (d) marks the region of 2nd peak in (a) at lower $T$ than Curie temperature where the principle ΔT peak takes place.

Both the $\Delta T$ and $\Delta S$ display similar changing patterns for $n$ = 2-10; however, their maxima gradually move to higher and lower temperatures (respectively) as $n$ decrease from 2 to 0 [see Figures SI2(c-d)]. In particular, the $n = 0$ $\Delta S$ behavior is shown in Figures 2(o) and SI3 and can be seen to monotonically increase with decreasing $t$ and to saturate at $t = 0$. The overall picture shown by Figures (2) and SI(2) implies that, compared with normal FEs, $\Delta T$ and $\Delta S$ enhancements can be expected for $n < 4.5$ and $n < 2$ EC materials. This suggests organic FEs, relaxor EC materials and tuning strategies of reducing transition diffuseness should be considered in the race to accelerate the realization of flexible and room-temperature EC cooling technologies.

Our model has also been applied to available $\Delta T$ and $\Delta S$ data from literature to further verify its versatility. $BaTiO_3$[11] and P(VDF-TrFE-CFE) 59.2/33.6/7.2 mol% terpolymer[30] film have been chosen as a representative normal FE and organic EC structures, respectively, and their directly-measured $\Delta T$ changes with $E$ are best fitted in Figure 3(a) adopting equation (2). Within our EC framework, the as-shown rapid $\Delta T$ saturation in $BaTiO_3$ ceramics follows from its long-range correlation character and hence a sizable $V_{cr}$ while the P(VDF-TrFE-CFE) polymer undergoes incomplete growth even at $E$=150 MV/m, confirming its ultrahigh breakdown field and extremely small $V_{cr}$. These inferences are consistent with our fitting result for $V_{cr}$ of 86.5 nm$^3$ for $BaTiO_3$ ceramics and $V_{cr}$=2.6 nm$^3$ for relaxor P(VDF-TrFE-CFE) terpolymer. Moreover, equation (2) has been applied to the $E$-dependence of $\Delta S$ in relaxor 0.65PMN-0.35$PbTiO_3$ (0.65PMN-0.35PT)[31], $Ba(Zr_{0.15}Ti_{0.75})O_3$ (BZT)[32] and $(Pb_{0.88}La_{0.08})(Zr_{0.65}Ti_{0.35})O_3$ (PLZT)[33] films, as shown in Figure 3(b). It can be seen that our curves tally with the experimental data, except that the BZT curve underestimates the entropy change at lower $E$. This slight discrepancy in BZT is probably caused by the strong $E$-dependence of its correlation volume - although we determine $V_{cr}$ = 71 nm$^3$ for BZT film (comparable to that of $BaTiO_3$ ceramics), there might also exist an $E$-induced AFE to FE phase transition in the BZT sample[32]. This suggests that such unique AFE to FE transitions need special consideration when calculating EC cooling properties of AFE materials, which is exemplified by $Pb_{0.95}Zr_{0.05}TiO_3$ (PZT) thin films with its direct $\Delta S$ data at 270 K and 330 K, shown in Figure 3(c). In particular, the stabilization of a long-range correlated FE phase is confirmed in the PZT film, formed out of an initial tiny AFE phase when $E$ exceeds a threshold ($E_{TP}$=53.3 and 58.8 MV/m at 330 K and 270K). The PZT $\Delta S$ curve therefore constitutes three parts - a pure AFE region, a region of coexisting AFE and FE phases and an $E$-stabilized FE phase region. The $V_{cr}$ is calculated as 29 $nm^3$ and 29.8 $nm^3$ for the FE phase at 330 K and 270 K, respectively, and as 3.2 $nm^3$ and 1.3 $nm^3$ for the AFE phase. The latter two are useful to explain the smaller $E_{TP}$ at 330 K, because it is well-known that stabilizing a long-range order parameter in a matrix of larger polarizable elements requires less energetic activation. Our results indicate that the $E$-stabilized FE phase in PZT thin film is nearly $T$-independent and, via the $V_{cr}$-dominated coefficient in equation (2), dominates the convergence of already saturated $\Delta S$ curves at ultrahigh $E$.

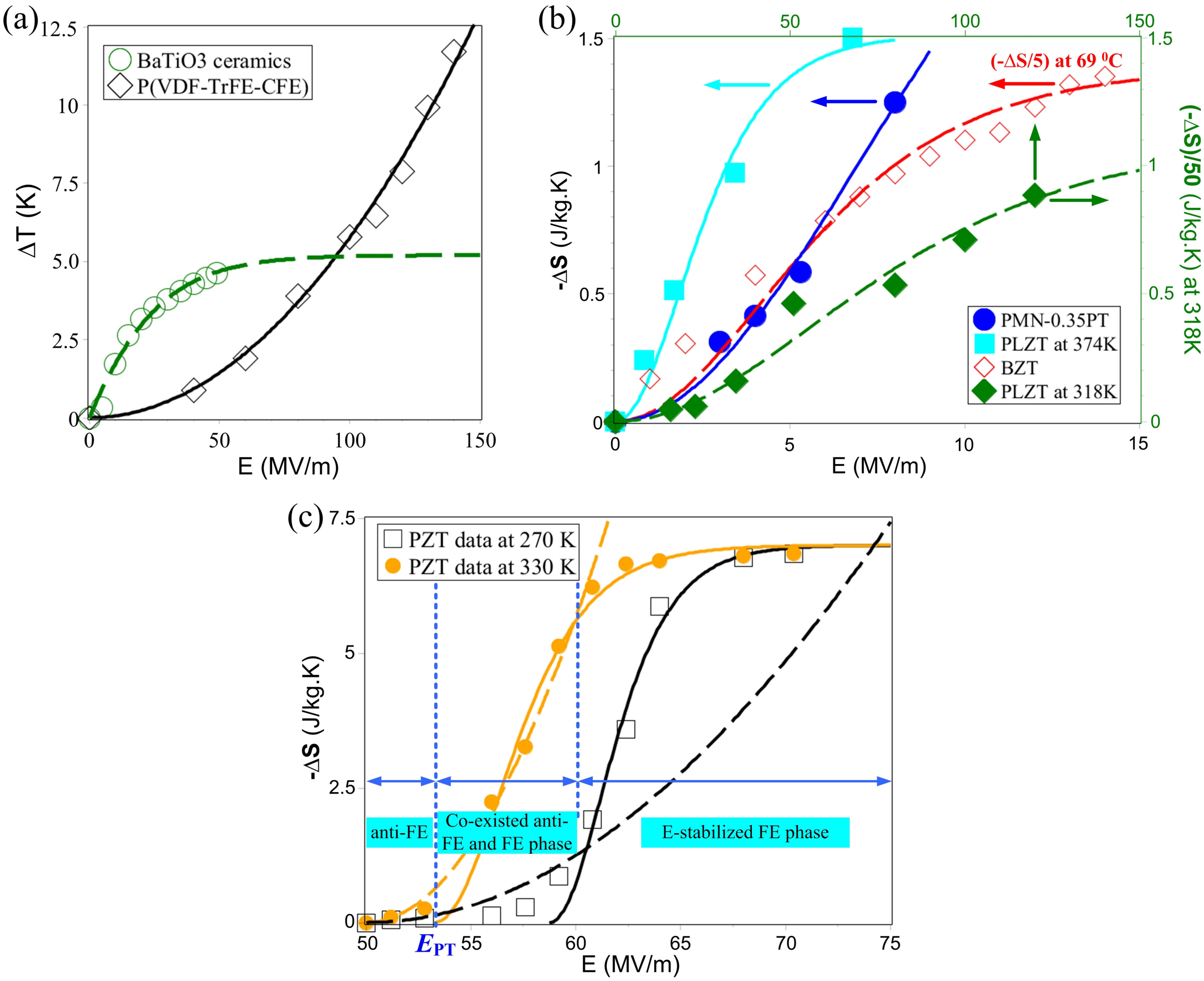

**Figure 3 | Theoretical predications and fittings for both ΔT and ΔS changes with $E$ in representative EC materials.** (a) Our fitting curves for experimental ΔT data in normal FE $BaTiO_3$ (open circles) and relaxor P(VDF-TrFE-CFE) 59.2/33.6/7.2 mol% terpolymer (open boxes). (b) Fittings of the ΔS data in 0.65PMN-0.35PT, BZT and PLZT. (c) Direct comparison of calculated (-ΔS) curves vs $E$ with two sets of indirect experimental data in PZT AFE thin film. The $BaTiO_3$, P(VDF-TrFE-CFE), 0.65PMN-0.35PT, BZT, PLZT and PZT data is derived from refs. 11, 30, 31, 32, 33 and 34, respectively.

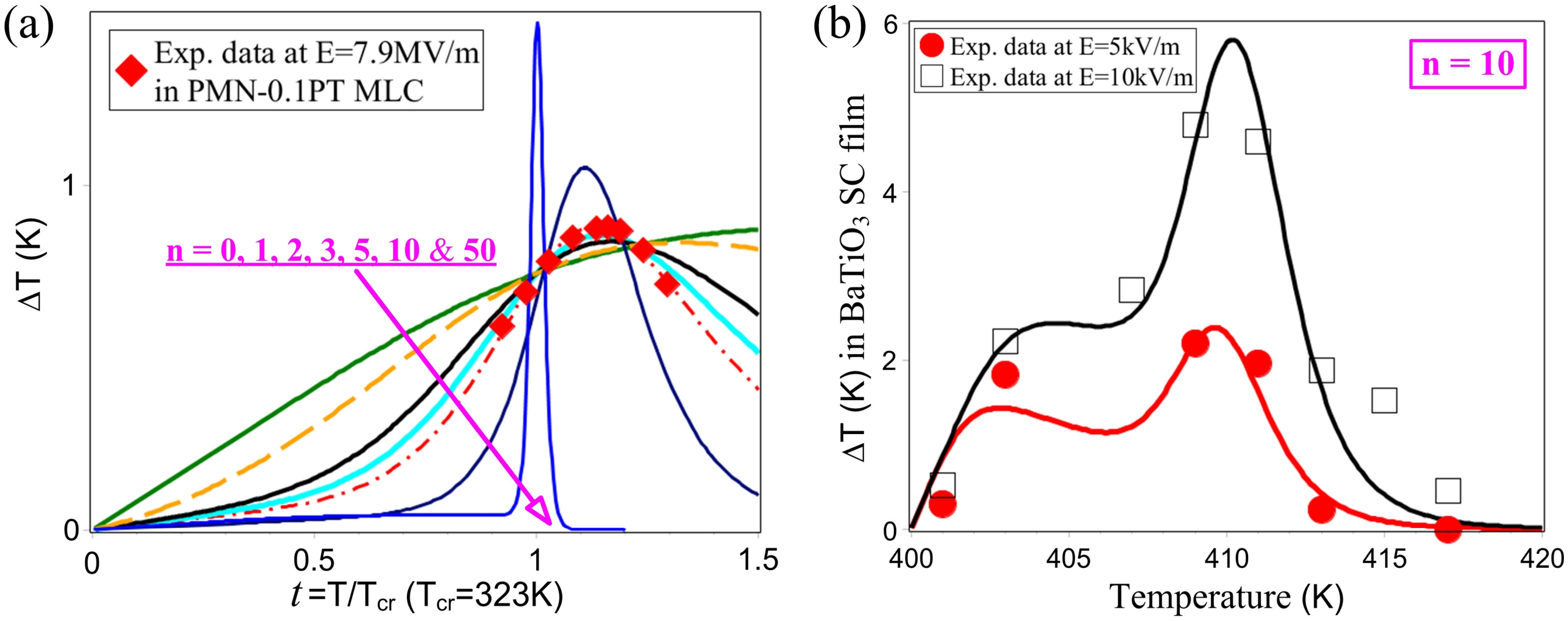

**Figure 4 |** (a) Calculated ΔT curves based on equation (2) with various *n* and correspondence to directly measured ΔT data in 0.9PMN-0.1PT MLCs at *E*=7.9 MV/m. (b) ΔT predications of Equation (2) using *n*=10 for two sets of directly measured ΔT data in $BaTiO_3$ single crystal (SC) film subjected to *E*=5 kV/m and 10 kV/m. The experimental data in (a) and (b) is derived from ref. 16 and ref. 11, respectively.

In addition to the proven versatility in quantifying the *E*-dependence of cooling responses in representative low-dimensional EC materials, equation (2) has been utilized to characterize EC responses as a function of temperature and contact with available data in literature. Directly measured $\Delta T$ data in normal FE $BaTiO_3$ single crystal films[11] and relaxor 0.9PMN-0.1PT multilayer capacitors (MLCs)[16] illustrated in Figure 4 serve as our database for our calculations of EC $\Delta T$ changes with *t*. The $\Delta T$ change in 0.9PMN-0.1PT MLCs shows a broad dispersion between 0.6 K and 0.9 K over a wide temperature range of 120K; whereas, subjected to $E$ = 10 kV/m, the $BaTiO_3$ $\Delta T$ rises steeply upto 4.8 K and abruptly decays to zero within a rather small temperature range of less than 5% $T_c$. These dissimilar thermal dispersions of $\Delta T$ can be considered as a natural consequence of distinct quasi-1$^{st}$ order phase transition in normal FE $BaTiO_3$ and the highly diffused depolarization dynamics in the relaxor 0.9PMN-0.1PT. Our $\Delta T$ calculations, on the basis of equation (2) with $n$=0, 1, 2, 3, 5 and 10, are plotted in Figure 4(a) for a direct comparison with the 0.9PMN-0.1PT data, which suggests that the *n* of 0.9PMN-0.1PT MLC lies between 2 to 5. The $\Delta T$ curve of an unrealistically high $n$ = 50 is also plotted for comparison and may be regarded as the limiting case that recovers the EC features of a 1$^{st}$ order transition. It can be seen that the $\Delta T$ vs. $t$ curves, as a whole, reproduce the major figures of merits of distinctly diffused EC materials - a larger $n$ essentially brings about a sharper but bigger EC peak close to $T_{cr}$; diffusive phase transitions are manifested in a broad EC peak dispersion, yet lower in magnitude and located far above $T_{cr}$. Our $\Delta T$ calculations yield $V_{cr}$=21 nm$^3$ for the 0.9PMN-0.1PT relaxor, which is appropriately larger than that in the PMN ceramics, $V_{cr}$=16 nm$^3$ determined from the ECE responsivity data[36] in Figure SI(4), since the normal FE nature of PT solid is expected to increase the correlation length and strength of relaxor 0.9PMN-0.1PT solid solutions.

In order to avoid meaninglessly high values of $n$, our $\Delta T$ calculations for the $BaTiO_3$ single crystal were restrained within a narrow $T$ range of 400 K to 420 K where noticeable $\Delta T$ changes are observed. In view of the lower $E$ (than the $E$ for the 0.9PMN-0.1PT MLC) which was applied to adiabatically depolarize the $BaTiO_3$ films, the novel implications of Figure 2(a) predict the possible occurrence of the dual peak phenomenon in the measured $BaTiO_3$ films. This can be used to explain the obvious shoulder in the $\Delta T$ change for $BaTiO_3$ at $E$=10 kV/m prior to the occurrence of the principle peak exactly at the Curie temperature and to clarify the absence of a clear maximum in $\Delta T$ at $E$=5 kV/m. We choose the $T_{cr}$ of $BaTiO_3$ as 410 K and 410.5 K for the $E$=5 kV/m and 10 kV/m data, respectively, while maintaining other parameters for the two $\Delta T$ datasets from $BaTiO_3$. A value of $V_{cr}$=108 nm$^3$ was determined for the $BaTiO_3$ single crystal film in Figure 4(b), which is larger than the value we of $V_{cr}$=86.5 nm$^3$ that we obtained for $BaTiO_3$ ceramics films in Figure 3(a). This expected $V_{cr}$ decease in ceramic samples, along with significantly improved breakdown fields, underlines the feasibility of ceramic processes in reducing the correlation length and broadening the electric de-poling scope, thus offering promising routes to enhancing the cooling properties of strongly-correlated oxide EC materials.

## Discussion

We present a versatile EC theory, based on a combination of a Master equation and Maxwell relations,

to analytically correlate the macroscopic EC cooling responses with sharp and diffusive phase transition characteristics. The adiabatic application of increased electric fields is found to trigger a quadratic increase of both $\Delta T$ and ($-\Delta S$). Subsequently, these quantities increase linearly with $E$ and ultimately saturate at high enough $E$. It worth noting that a steep decline of the change in $\Delta T$ with $E$ from 1 to 0.6 has been recently been reported near the morphotropic phase boundary in doped lead-free $(Na_{0.5}Ba_{0.5})TiO_3$-PT ceramic[9] EC films. Closely examining equation (2) allows one to approximate the final saturation trend as $\sim\left(-2u/e^{2u}\right)$, which further leads to an exponential decay of ECE responsivity as $E$ increases. The latter, together with the initial quadratic and then linear increases of the $\Delta T$ change, gives an overall picture of the $E$-dependence of ECE responsivity, which may be used to better understand the ECE responsivity data found in literature [see Figure SI(4)]. Also, the proportionality of the ECE responsivity maxima to $P^m$ requires piecewise functions for AFEs such that the $E$-induced AFE to FE phase transition and resultant $P^m$ increase be included to represent typical double $P$-$E$ hysteresis loops[4] in AFE EC materials.

Based on the correlation volume results from our EC calculations and theoretical fittings for a rich range of representative EC low dimensional structures, we are able to determine the transition diffuseness index of these EC materials and our $n$ results are shown in Figure 5(a). Universal linearity between $V_{cr}$ and $n$ is evident. It is striking yet critical to develop this generic linearity since it could directly explain the divergent sharpness and diffuseness of observed EC responses to localized correlation volume and activation energies for adiabatic depolarizations. It is also worth noting that such a functional bridge can span over substantial length scales, exemplified by the calculated values of $V_{cr}$=1.9 nm$^3$ and 1.6 nm$^3$ for P(VDF-TrFE) copolymer and $Ba(Zr_{0.15}Ti_{0.75})O_3$, to as high as ~100 nm$^3$ for $BaTiO_3$. In view of the fact that the $BaTiO_3$ $V_{cr}$ acts merely at its FE domain walls, which separate FE domains with much longer dimensions and effectively mediate $E$-induced polarization switching, it is therefore useful to postulate that the $n$-$V_{cr}$ linear correlation bridges EC studies from micrometer domain transition scale down to unit cell repeat scales. However, within such a large spread across such broad length scales, the dimensionality of EC phase transitions and the surrounding elastic and electronic conditions are not expected to be identical, and this in turn significantly alters the EC cooling behaviors.

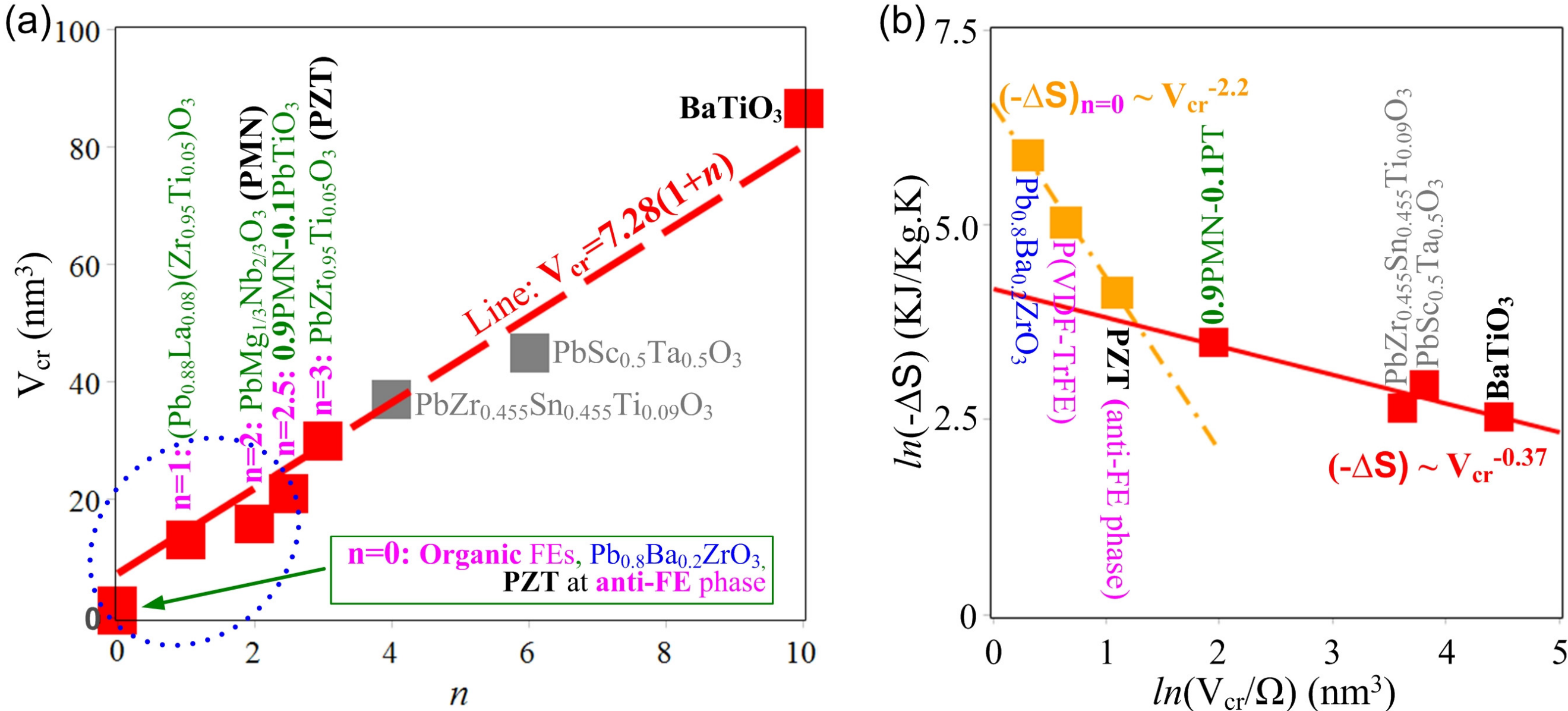

**Figure 5 | A linear dependence of correlation volume on transition diffuseness index and Distinct scaling laws of ΔS in typical EC materials.** (a) A linear relation between $V_{cr}$ and $n$ in typical EC materials. (b) Distinct scaling laws of ΔS in $n$>2 EC materials exemplified by $BaTiO_3$ normal FE and relaxor FEs and in $n$=0 EC materials including $Pb_{0.8}Ba_{0.2}ZrO_3$ oxide, AFE $Pb_{0.95}Zr_{0.05}TiO_3$ and FE P(VDF-TrFE) 65/35 mol.% copolymer. The symmetry factor Ω is 3 for 0.9PMN-0.1PT and 1 for others. The dotted ellipse encloses the EC materials expected to possess significantly enhanced entropy change. The ΔS data in (b) is derived from ref. 1.

The effects of EC transition dimensionality on Δ*S* scaling behaviors are distinguished in Figure 5(b). It can be seen Δ*S* in *n*=0 EC materials, including polymeric FEs, AFE phase $Pb_{0.95}Zr_{0.05}TiO_3$ and $Pb_{0.8}Ba_{0.2}ZrO_3$, scales proportionally with $V_{cr}^{-2.2}$; whereas Δ*S* in $Pb_{0.95}Zr_{0.05}TiO_3$, relaxor and normal FE EC materials scales distinctively as $\Delta S \sim V_{cr}^{-0.37}$. Recalling the extremely small (< 3 $nm^3$) correlation volume in the *n*=0 EC materials and the recently-proposed inversely quadratic relation of Δ*S* with correlation length[37], $\Delta S \sim V_{cr}^{-2.2}$ is fundamentally ascribed to one-dimensional structural confinements of lattice-scale depolarization dynamics in the *n*=0 EC materials and structures and as well as in the remaining two degrees of freedom of concomitant entropy changes. These novel findings suggest that the upper bound of EC entropy change can be limited by the unit cell volume. Thus replacing the $V_{cr}$ in our derived $(-\Delta S)_{max}=\ln(2)\Omega k/V_{cr}$ with unit cell volumes reported in ref. 27 permits us to predicate values for $(-\Delta S)_{max}$ as high as 300 kJ/(K.m$^3$) and 215 kJ/(K.m$^3$) for $Pb_{0.8}Ba_{0.2}ZrO_3$ and P($VDF_{0.65}TrFE_{0.35}$), respectively. In contrast, the distinct scaling of $\Delta S_{max} \sim V_{cr}^{-0.37}$ in rich relaxor and normal FE EC materials tallies with the theoretically demonstrated long-range dipolar correlation exponent of 2/3 in rough polar interfaces[38]. The well-known FE domain wall dimensionality of 2.5 and fractional dimensionality of PNR evolutional dynamics in relaxors[21,26] are also essentially responsible for the $\Delta S_{max} \sim V_{cr}^{-0.37}$ scaling.

Our findings provide guidance for characterizing known low-dimensional EC materials and designing new ones with enhanced cooling performance by deliberately increasing phase transition diffusivity and shortening the correlation length. Following this logic, the EC enhancement in polymeric FEs and AFEs have much higher sensitivity to nanoconfinement and truncation in the characteristic correlation volume. The distinct scaling laws of maximum EC entropy change have potential applications in detecting new minority phases in phase-coexisting EC materials and for extrapolating electric and thermal activations for which current EC cooling data are not available yet. Also, our work may have critical implications beyond EC cooling properties, as an analogous theoretical framework could be established for other extensively-studied caloric effects such as magnetocalolric and mechanocaloric effects.

## Methods

The Master equations are introduced to derive a generic expression for the macroscopic polarization in EC materials as a function of thermal and electric activations. The defined microscopic polar elements (MPEs) are assumed to be embedded within a paraelectric matrix, but mutually correlated at a variety of length scales. At sufficiently high $T$ and zero $E$, the localized dipoles in EC MPEs are randomly oriented, cancelling out the overall polarization for ECE. When activated by thermal cooling or $E$-poling along the axis of lattice symmetry, the polarization of partial MPEs reorientates and aligns along the closest lattice-symmetry allowed direction, leading to the emergence and increase of overall polarization magnitude and hence entropy decrease. The thermal cooling and $E$-activated partial alignments of localized polarization among all the MPEs lead to an equivalent number density of MPEs occupying a favored ground state. The time ($t$)-dependent probability ($p_n$) of the $n^{th}$ ground state occupied by EC MPEs is governed by the Master equations:

$$\tau \frac{d}{dt} p_n(t) = \sum_{m \neq n} \left[ p_m(t) - e^{\mu(n,m)} p_n(t) \right]$$

where integer $m \neq n$, $\tau$ denotes characteristic relaxation time and $\mu(n \leftrightarrow m)$ is the energy difference between the $m^{th}$ and $n^{th}$ state. For an EC material with two ground states, $\mu(0,1)=Q\mathrm{V}_{MPE}/kT$ with $k$ Boltzmann's constant. Here $Q$ corresponds to the activation energy density given by $2E \times P^m$ with $P^m$ denoting an $E$- and $T$-independent polarization maxima attainable in the two-state system. At high limit and under $E$-poling applied along the axis of lattice symmetry, the Master equations give a generic expression of the macroscopic polarization magnitude ($P$):

$$P = P^m \tanh\left( EP^m \mathrm{V}_{MPE} / \Omega kT \right)$$

Where $u(E,T) = EP^m \mathrm{V}_{MPE} / (\Omega kT)$ represents a universal activation parameter of both $E$ and $T$; $\Omega$ is a symmetry contribution factor equaling to 1 and 3 for a two-state (e.g. tetragonal and polymeric) and eight-state (e.g. rhombohedral) EC material, respectively.

## Acknowledgements

Y.S. gratefully acknowledges the Research Grants Council of Hong Kong SAR for Hong Kong Fellowship funding; L.H. and Y.S. are supported by the Basic Research Fund of Shenzhen (Grant No. JCYJ20150507170334573); A.K.S. and Y.S. acknowledge the FRGS Grant (Project No.

FRGS/2/2013/SG06/MUSM/01/1) provided by the Ministry of Higher Education, Malaysia, and the Advanced Engineering Programme and School of Engineering, Monash University Malaysia; G.J.W. is supported by US National Science Foundation (Grant No. CMMI-1162431); S.L. thanks the Hundreds talents program of Chinese Academy of Sciences (Grant No. R52A261Z10) and the Fundamental and advanced technology Research Funds of Chongqing (Grant No. cstc2015jcyjbx0103); S.A.T.R. is grateful for support from the British Council Newton Fund (Grant No. 172724105).

## Author contributions

Y.S. conceived the concept and framework. Y.S. derived the theory and prepared the figures while all authors analyzed the results. Y.S., L.H. and S.A.T.R wrote the manuscript. All authors reviewed and edited the whole paper.

## Additional information

**Supplementary information** accompanies this paper.

**Competing financial interests:** The authors declare no competing financial interests.

# Supplementary Information to:

# Distinct scaling behaviors of giant electrocaloric cooling performance in low-dimensional organic, relaxor and anti-ferroelectrics

**Yuping Shi[1,2,3], Limin Huang[2,*], Ai Kah Soh[3], George J. Weng[4], Shuangyi Liu[5] & Simon A.T. Redfern[6,7]**

[1]Department of Mechanical and Aerospace Engineering, Hong Kong University of Science and Technology, Clear Water Bay, Kowloon, Hong Kong.
[2]Department of Chemistry, South University of Science and Technology of China, Shenzhen 518005, China.
[3]School of Engineering, Monash University Malaysia, Bandar Sunway 46150, Malaysia.
[4]Department of Mechanical and Aerospace Engineering, Rutgers University, New Brunswick, New Jersey 08903, USA.
[5]Chongqing Institute of Green & Intelligent Technology, Chinese Academy of Sciences, Chongqing 400714, China.
[6]Department of Earth Sciences, University of Cambridge, Downing Street, Cambridge CB2 3EQ, UK.
[7]HPSTAR, 1690 Cailun Rd, Pudong District, Shanghai 201203, China.

[*]Correspondence and requests for materials should be addressed to L.H. (email: huanglm@sustc.edu.cn)

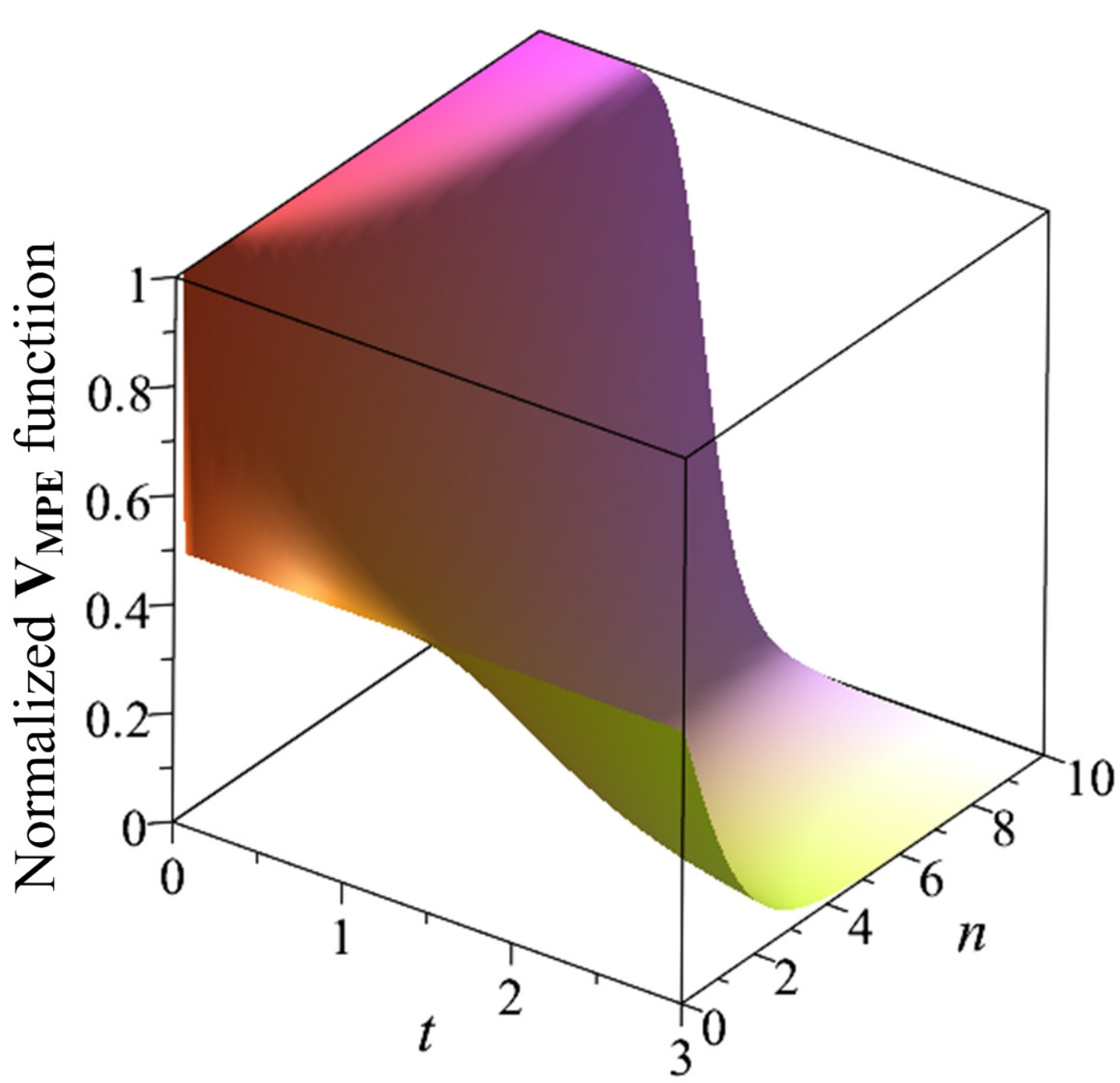

**Figure SI1.** A 3D plot of differently diffused MPE volume (normalized by $V_{cr}$) in EC materials as a function both $t$ and $n$.

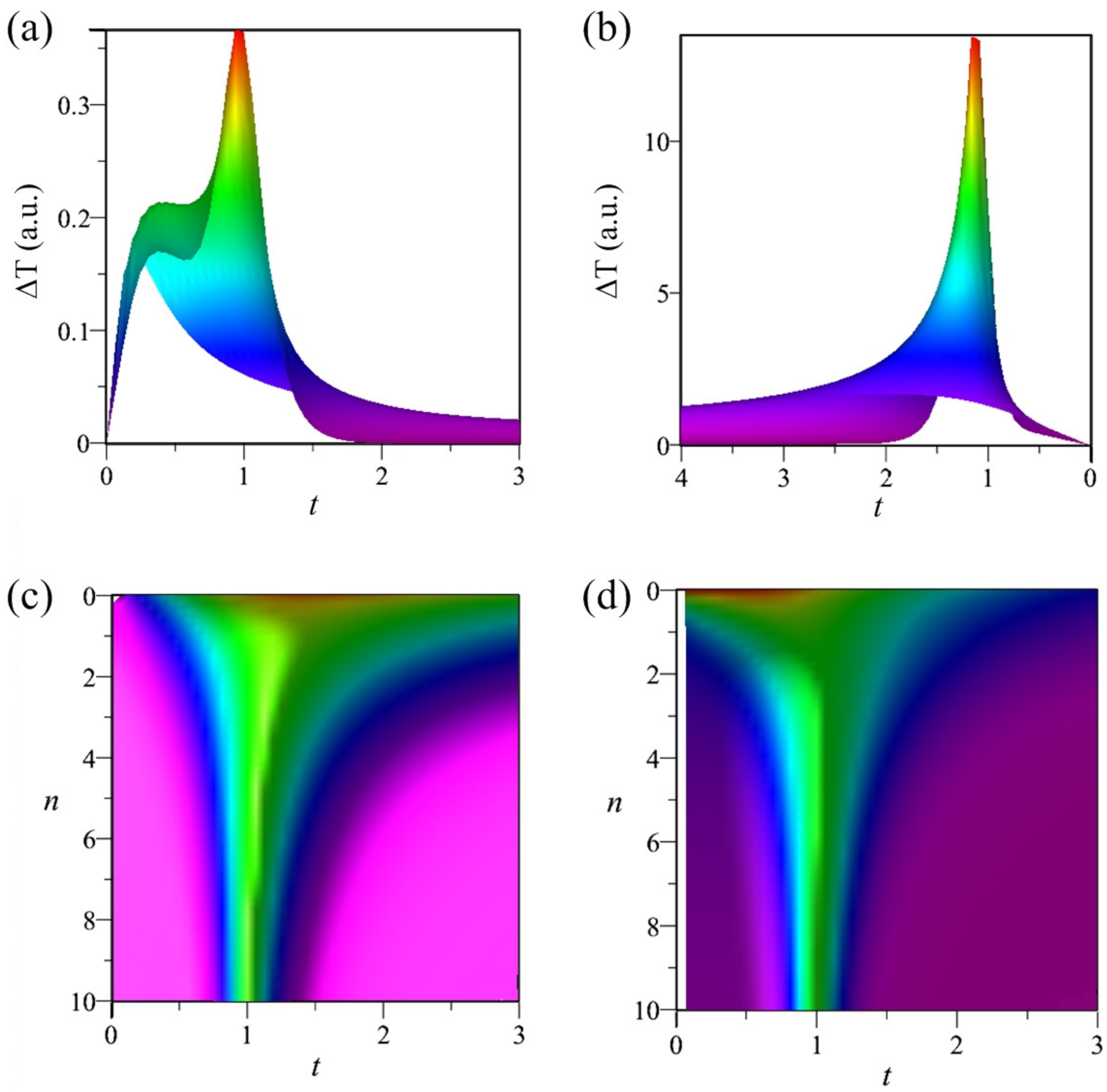

**Figure SI2.** (a, b) ΔT-*t* plane projection of calculated ΔT 3D surfaces in Figures 2(a) and 2(b) in the main paper; (c, d) *n*-*t* plane projection of the ΔT and (-ΔS) 3D surfaces shown in Figures 2(d) and 2(e).

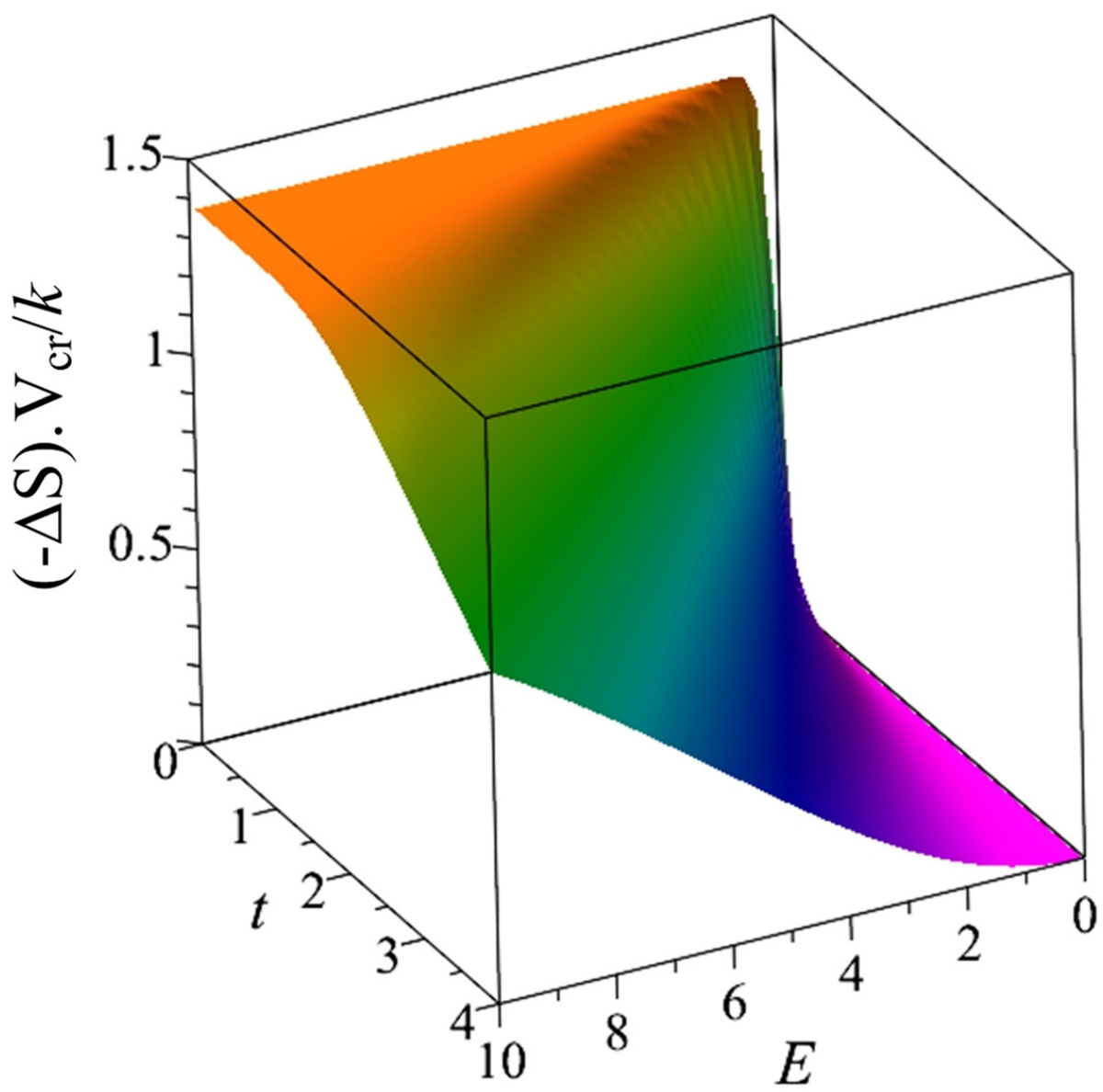

**Figure SI3.** A 3D plot of negative entropy change (normalized by $k/V_{cr}$) as a function of both $t$ and $E$ activations in the $n$=0 electrocaloric materials and structures.

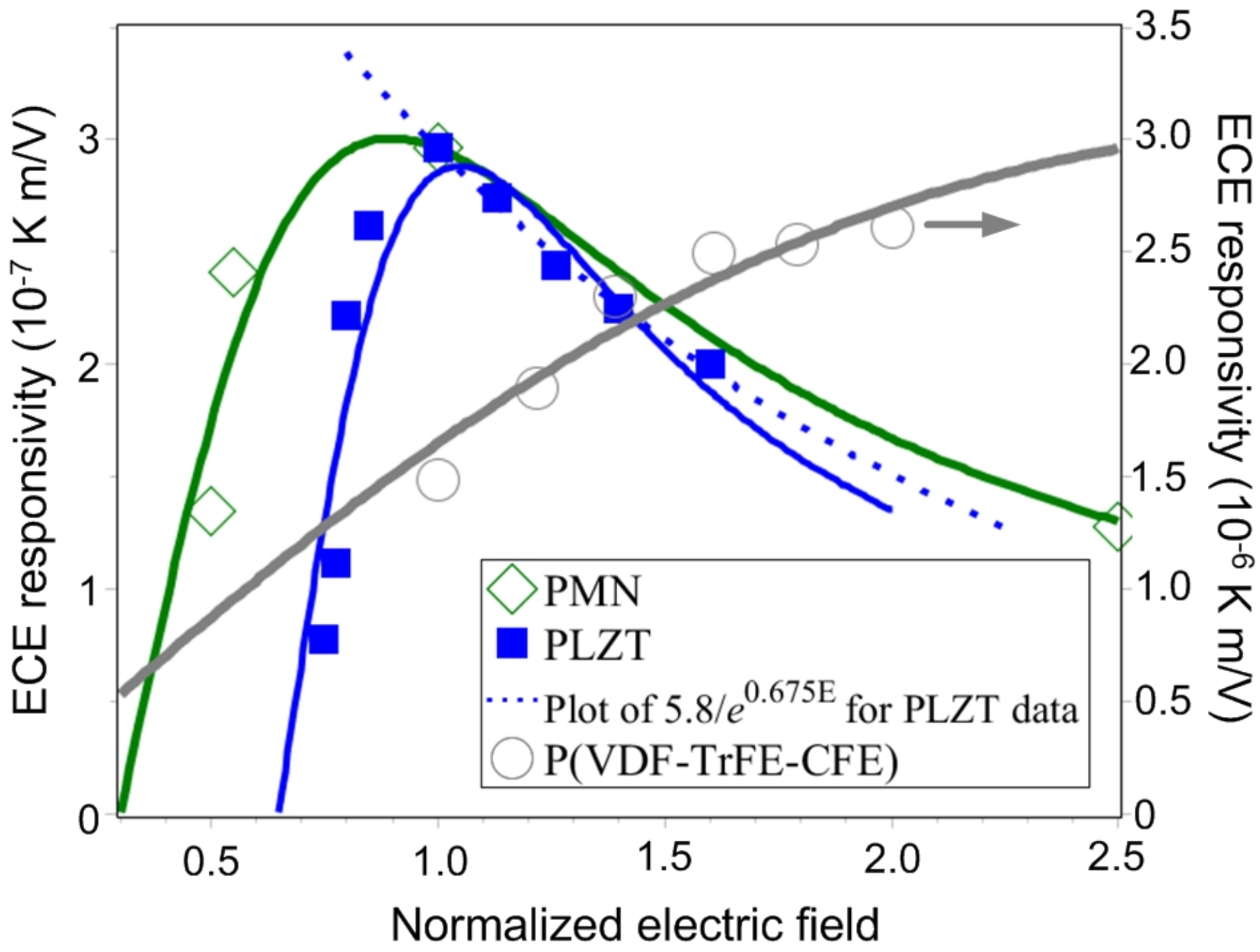

**Figure SI4.** Our ECE responsivity fitting curves for the experimental data measured in PMN, PLZT and P(VDF-TrFE-CFE). An exponential decay is also plotted for the PLZT data for a direct comparison. The PMN and PLZT data are derived from ref. 36; P(VDF-TrFE-CFE) data are from ref. 10. Reference codes and material abbreviations refer to the main paper.